\documentclass[pdflatex,sn-mathphys]{sn-jnl}

\usepackage{adjustbox}
\usepackage{subcaption}
\usepackage{float}
\usepackage{xr}
\usepackage{hyperref}

\makeatletter

\jyear{2023}%

\theoremstyle{thmstyleone}%
\theoremstyle{thmstyletwo}%

\theoremstyle{thmstylethree}%

\begin{document}

\title[Reconstruction of Renal Arterial Network]{A Hybrid Approach to Full-Scale Reconstruction of Renal Arterial Network}


\author*[1]{\fnm{Peidi} \sur{Xu}}\email{peidi@di.ku.dk}

\author[2]{\fnm{Niels-Henrik} \sur{Holstein-Rathlou}}\email{nhhr@sund.ku.dk}

\author[2]{\fnm{Stinne Byrholdt} \sur{Søgaard}}\email{sbyrholdt@sund.ku.dk}
\author[3]{\fnm{Carsten} \sur{Gundlach}}\email{cagu@fysik.dtu.dk}

\author[2]{\fnm{Charlotte Mehlin} \sur{Sørensen}}\email{cmehlin@sund.ku.dk}

\author[1]{\fnm{Kenny} \sur{Erleben}}\email{kenny@di.ku.dk}

\author[2]{\fnm{Olga} \sur{Sosnovtseva}}\email{olga@sund.ku.dk}
\equalcont{These authors contributed equally to this work.}

\author[1]{\fnm{Sune} \sur{Darkner}}\email{darkner@di.ku.dk}
\equalcont{These authors contributed equally to this work.}

\affil[1]{\orgdiv{Department of Computer Science}, \orgname{University of Copenhagen}, \orgaddress{\street{Universitetsparken 1}, \city{Copenhagen}, \postcode{2100}, 
\country{Denmark}}}

\affil[2]{\orgdiv{Department of Biomedical Sciences}, \orgname{University of Copenhagen}, \orgaddress{\street{Blegdamsvej 3B}, \city{Copenhagen}, \postcode{2200}, 
\country{Denmark}}}


\affil[3]{\orgdiv{Department of Physics}, \orgname{Technical University of Denmark}, \orgaddress{\street{Kongens Lyngby}, \city{Copenhagen}, \postcode{2800}, 
\country{Denmark}}}












\abstract{The renal vasculature, acting as a resource distribution network, plays an important role in both the physiology and pathophysiology of the kidney. However, no imaging techniques allow an assessment of the structure and function of the renal vasculature due to limited spatial and temporal resolution.
To develop realistic computer simulations of renal function, and to develop new image-based diagnostic methods based on artificial intelligence, it is necessary to have a realistic full-scale model of the renal vasculature. 
We propose a hybrid framework to build subject-specific models of the renal vascular network by using semi-automated segmentation of large arteries and estimation of cortex area from a micro-CT scan as a starting point, and by adopting the Global Constructive Optimization algorithm for generating smaller vessels.
Our results show a statistical correspondence between the reconstructed data and existing anatomical data obtained from a rat kidney with respect to morphometric and hemodynamic parameters.}




\maketitle

\section{Introduction}
\subsection{Biological background}
In each organ, the vasculature has a characteristic structure adapted to meet the specific needs of the organ. In the kidney, the vasculature plays a special role. Not only does it function as a resource distribution network, supplying the individual nephrons with blood and nutrients, but it also constitutes a communication network, allowing contiguous nephrons to interact through electric signaling along the vessels \cite{marsh2019nephron}.

Nordsletten et al. \cite{nordsletten} provided the hitherto most detailed and quantitative description of the rat renal vasculature. They combined high ($4\ \mu m$) and low ($20\ \mu m$) resolution micro-CT images obtained from a vascular cast of a rat kidney. They used the skeletonization method to trace the path of contiguous vessels and then applied the Strahler approach \cite{strahler1952topology} to sort and interpret the data. Strahler ordering sorts treelike networks by the diameter of the branches according to a bifurcating scheme. The principal assumption required for its use is the existence of a diameter-based hierarchy of vessels, ending in the narrowest vessels, i.e., the afferent arterioles supplying the individual nephrons.

Marsh et al.  \cite{marsh2017network} used micro-CT with $2.5\ \mu m$ resolution to assess the three-dimensional microvascular structure of the rat renal arterial tree. The cast revealed an arterial tree network originating in arcuate arteries, branching as few as twice or as many as six times before reaching a terminal artery that terminated in pairs, triplets, or quadruplets of afferent arterioles. Marsh et al. identified different motives for how afferent arterioles originated from all branch orders of nonterminal arteries or from terminal arteries forming the tops of the arterial trees. Similar branching patterns have been reported by other groups using microdissected trees from four different mammalian species \cite{more1951vascular, horacek1987vascular,casellas1994vascular}.

Postnov et al. \cite{postnov2016renal} have shown that the pressure drop in a simple bifurcating tree with the vessel dimensions reported by Norsletten et al. \cite{nordsletten} exceeds the value found experimentally. This is expected since in a simple bifurcating tree afferent arterioles appear only at the terminal branch points of the tree – an assumption that maximizes the hemodynamic resistance between the renal artery and the glomerulus. Postnov et al \cite{postnov2016renal}  and Marsh et al. \cite{marsh2017network} have reported an exponential distribution of the distances between branch points for afferent arterioles across the vascular tree. This distribution, and the possibility to branch from any arterial segment, is the basis for the pressure in the glomerular capillaries being significantly higher, and in a range compatible with normal nephron function, than in a simple bifurcating tree. 

The number of nephrons in a kidney in a given species is variable and is thought to play an important role in renal health. Baldelomar et al. estimated nephron numbers from in vivo images and from high-resolution ex vivo images \cite{baldelomar2018nephrons}, while Letts et al. \cite{letts2017nephrons} assessed the location of glomeruli in the outer 30\% of the cortex,  midcortical nephrons (30 – 60\%), and juxtamedullary nephrons of the inner 40\% of the cortex. Both studies show high variability in the number and characteristics of nephrons. Taken together, the high variability, both in the structure of the renal vasculature and in the number of nephrons in a given kidney, suggests that a probabilistic-based approach to model nephron-vascular architecture and blood flow dynamics is the right choice.


\subsection{Modeling outlook}
Three major methods have been described in the literature to construct models of vascular networks. Pure rule-based models \cite{postnov2016renal,marsh2013multinephron,marsh2019nephron} generate vascular trees analytically from a given root while completely ignoring the spatial structure of the network. The length of each vessel, the radius distribution to its children in a bifurcation, as well as the stopping criteria are all derived from given probability distributions obtained from experimental data. Although the hemodynamics can be simulated without information on the spatial structure, these methods cannot generate real-looking networks and ignore the subject-specific information, and thus cannot be utilized for individual analysis. 

The image-based reconstruction methods build 3-D geometric models that capture the high-level structure of an individual’s blood vessels from clinical images \cite{he2020learning_segment,dorobantiu2021coronary_segment,schneider2015joint_vessel_segmentation}. These methods involve either a segmentation followed by a centerline extraction or a direct tracking of the blood vessels. Despite advances in Convolution Neural Networks, learning from very thin structures is still challenging and will suffer from errors due to both merging and discontinuity, resulting in extremely intense manual work afterward. More importantly, in the kidney the vessels at far-surface regions are beyond the experimental resolution, making it impossible to detect the small vessels from an image alone. These small vessels, however, are the ones supplying the individual nephrons and thus have to be resolved in the final model. Therefore, image-based reconstruction alone is unable to provide complete and detailed 3-D vasculatures in the kidney, making biosimulation the only tool available with generative models that can extrapolate modeling to unresolved parts of the kidney. 

The angiogenesis-based methods simulate the growth of vasculatures by considering the biological and physiological factors involved in the process. These algorithms model the vascular tree growing as an optimization problem following the assumption that 
the network achieves a topological and geometrical structure over the vascularized tissues from hemodynamic principles \cite{cury2021parallel}. There are two main methods to generate the vasculature based on this growing algorithm, namely, Constraint Constructive Optimization (CCO) method proposed by Schreiner and Buxbaum \cite{schreiner2000constrained}, and its variant Global Constructive Optimization (GCO) proposed by Georg et al. \cite{georg2010global}. Both the CCO and GCO algorithms grow the tree inside a pre-defined perfusion territory. In both types of algorithms, a single tree root location of the blood inlet is chosen manually. Besides, boundary conditions such as terminal radius and flow distributions are imposed to represent physiologic conditions. 

These methods are able to generate real-looking vascular structures with both spatial location and connectivity information and have been applied in the liver, heart (left ventricle), and eye  \cite{schreiner2000constrained,georg2010global,jaquet2018hybrid_ventricle}. Recently, Shen et al. \cite{shen2021mathematical} and Ii et al. \cite{ii2020multiscale} incorporated GCO and CCO, respectively, to reconstruct vasculatures in the human brain. Although Cury et al. \cite{cury2021parallel} has recently applied an adaptive CCO on a prototypical human kidney model, no similar research exists on real renal vasculatures due to the complex non-convex geometry. Moreover, most of the studies produce homogeneous vascular network models that do not account for individual differences, so they cannot be used for individual analysis.

Both CCO and GCO require a convex structure since the connections between any two nodes should not leave the structure. This is one of the reasons why it is challenging to adopt these methods to an organ like the kidney with a complex internal structure. Some parts, like the renal pelvis and the pyramids, also pose intrinsic spatial restrictions on vessel construction, which are difficult to model when the tree grows from only a single root node. 

Our work follows a similar idea of \cite{shen2021mathematical} by proposing a hybrid way to incorporate subject-specific image-based priors via a semi-automated segmentation of the main (large) arteries and an automatic cortex approximation from the ex-vivo micro-CT scan of a real rat kidney, which both are utilized in the GCO initialization step. Briefly, a prebuilt arterial tree consisting of main arteries is extracted from the main artery segmentation, while we also propose a novel approach to sample terminal nodes (glomerulus) from the estimated renal cortex while maximizing the distance between any two neighboring nodes using Poisson disk sampling \cite{bridson2007fast_poisson_sample}. 

These sampled terminal nodes are then connected to the pre-built vascular tree. Instead of growing from a single root position, the algorithm can now start growing from a pre-built vascular tree and thus will retain subject-specific information in the final generated full-scale vascular tree. At the same time, this procedure avoids violating the intrinsic spatial constraints, thus making the complex structure piece-wise convex. In contrast to \cite{shen2021mathematical} which proposes forest growth, our algorithm consists of a single tree but with pre-built large branches because of the single inlet of the renal artery. Our results show that the structural and functional properties of the reconstructed vascular network are in good agreement with existing anatomical data.


Currently, the image prior starts with a semi-automated segmentation of the vessels, which is time-consuming work. However, recent advances in convolutional neural networks are likely to be able to segment large vessels given a decent amount of labeled training data, especially in the case of micro-CT with relatively low variation among the images. Apart from the segmentation retrieval and a single user-defined root position, our work is fully automatic, meaning that no software interface is involved in the process. In particular, the leaf nodes sampling and centerline extraction are both implemented in pure Python, unlike \cite{shen2021mathematical} which used BrainSuite for cortex extraction, and \cite{ii2020multiscale} which used Amira for brain hemispheres extraction and skeletonization. All the 3-D software packages, e.g. 3D Slicer \cite{Kikinis2014_3dSlicer} and ParaView \cite{ahrens2005paraview} are only used for visualization.

\section{Results}
\subsection{Model implementation and rendering}
\begin{figure}[htb]
\begin{minipage}[b]{1.0\linewidth}
  \centering
  \centerline{
  \includegraphics[width=1.0\linewidth]{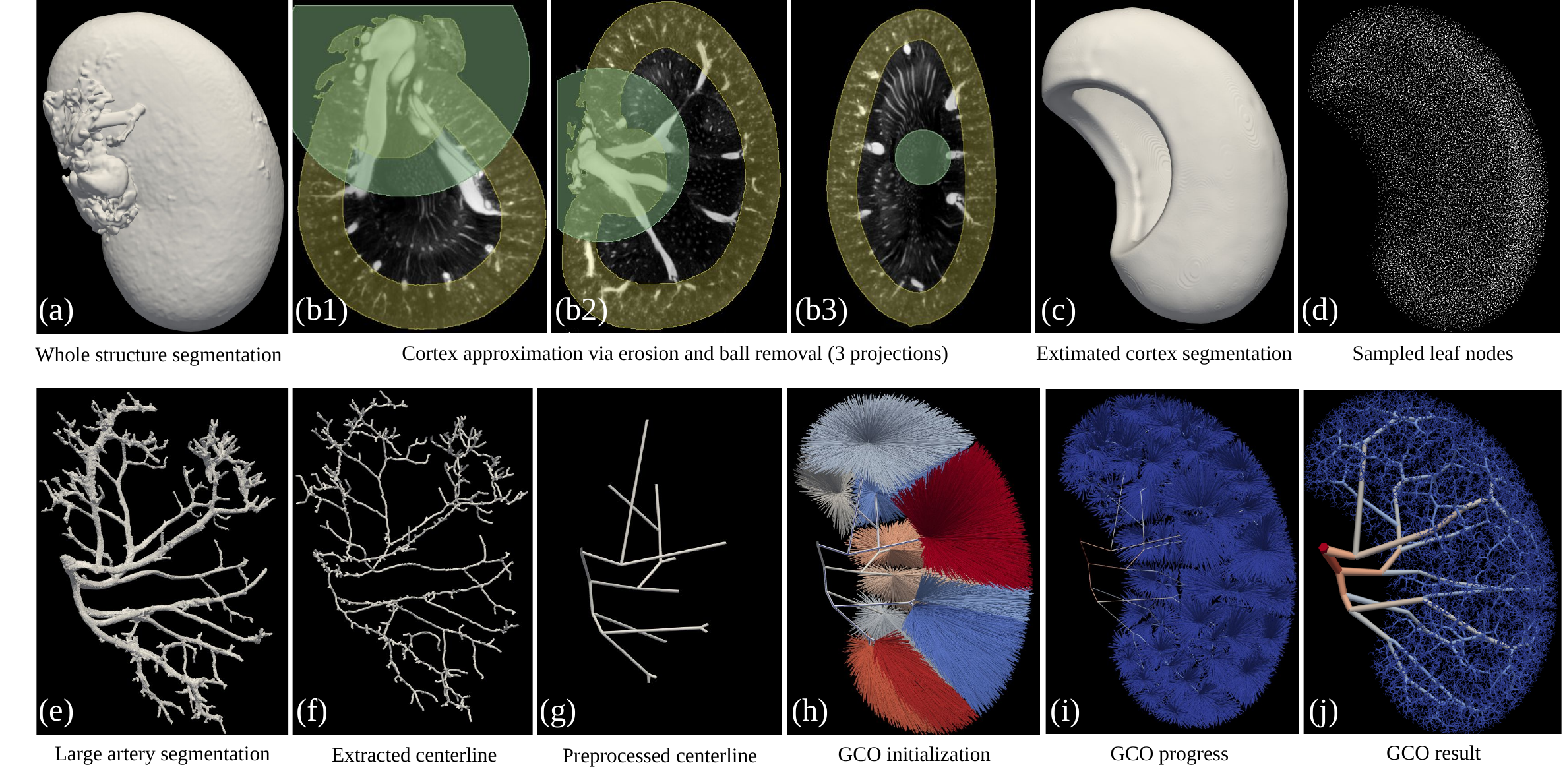}
  }
\end{minipage}
\caption{GCO Pipeline. The initial micro-CT scan is used to extract whole structure segmentation (a) and large artery segmentation (e). Top row: renal cortex (c) is approximated by a subtraction of erosion followed by a ball removal (b), where the leaf nodes (d) are sampled using Poisson disk sampling. Bottom row: extracted centerline (f) is pre-processed to pre-build a renal arterial tree consisting of only the first few large arteries (g). In GCO initialization (h), all the sampled leaf nodes (d) are connected to the nearest node in the pre-built tree (g) with color indicating the group of leaf nodes that are connected to the same node. Colors in the GCO progress and result (i\&j) indicate the radius of each vessel: from $300 \ \mu m$ in renal artery to $10 \ \mu m$ in afferent arterioles (AA).}
\label{fig:whole_process}
\end{figure}

Our hybrid modeling approach to reconstruct a renal vascular network combines semi-automated segmentation of large arteries from micro-CT images and the Global Constructive Optimization algorithm for the generation of smaller microvessels (Fig.~\ref{fig:whole_process} and is described in detail in Methods). The raw scan has an isotropic voxel size of $22.6 \ \mu m^3$. From Table 2 in Nordsletten's paper \cite{nordsletten}, renal arteries with Strahler order from 0 to 2 have a mean radius of $10.08$, $13.90$, $20.06 \ \mu m $ respectively, making it impossible to detect those small arteries from the $22.6 \ \mu m^3$ scans. Using our hybrid approach, the small arteries are successfully resolved, producing a full-scale renal arterial tree.

To demonstrate the validity of our reconstructed vascular tree from GCO, we visualize a reconstructed tree in 
Fig.~\ref{fig:whole_process}j. Each vessel is visualized by a separate cylinder with a thickness corresponding to its radius and colored coded by the radius. This gives us a first visual inspection of the realistic features of our generated vascular tree. 3D gif animations with rotation are available at: \url{https://github.com/KidneyAnonymous/RenalArterialTree}


\subsection{Morphometric validation}

\begin{figure}[H]
\centering
   \includegraphics[width=.7\textwidth]{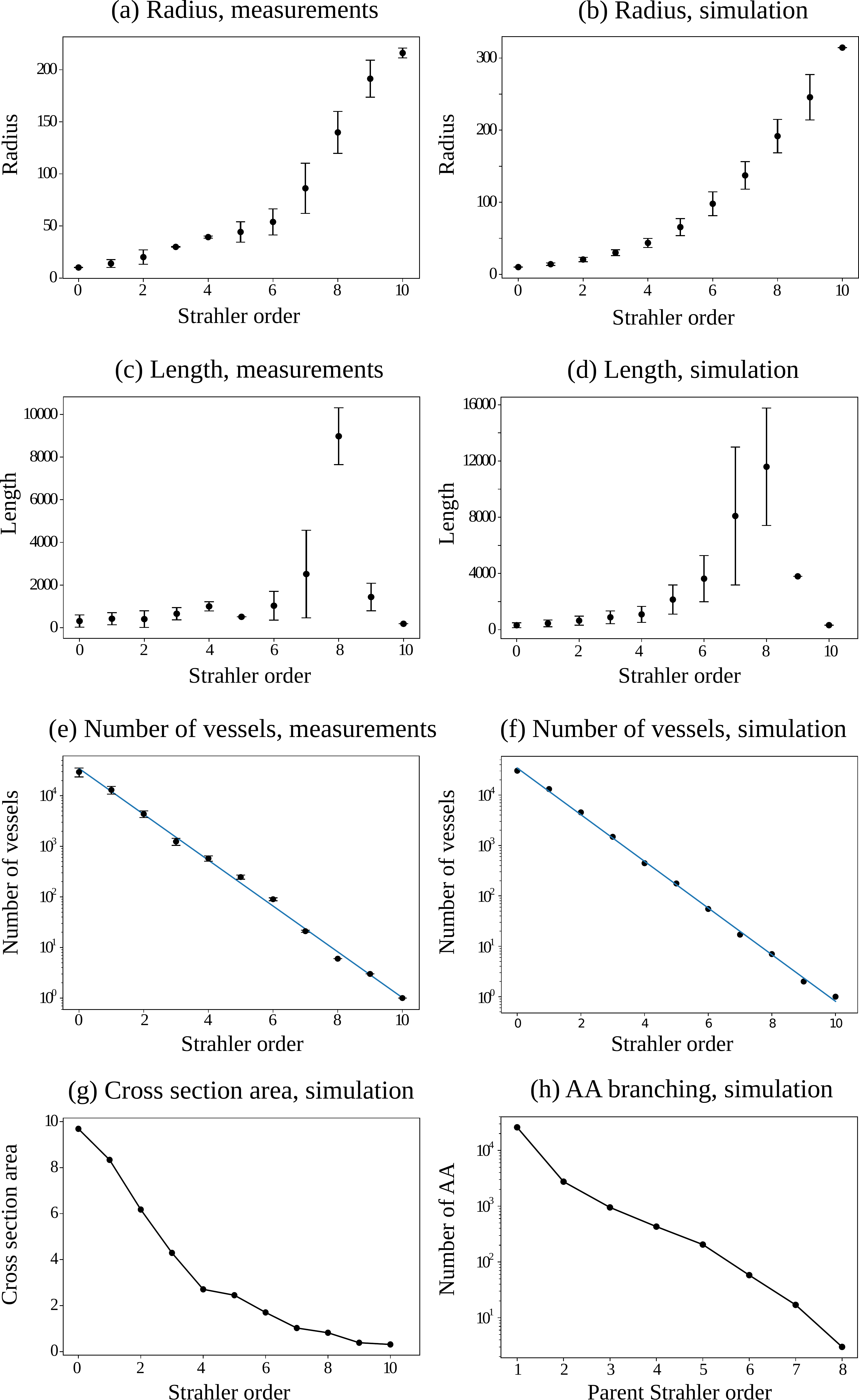}
 \caption{Morphometric features of the generated renal vascular arterial network. It shows good agreement with physiological measurements reported in the literature. (a) and (b) Vessel radius (in $\mu m$) vs Strahler order drawn from Ref.~\cite{nordsletten} (Table 2) and from our generated tree, respectively. (c) and (d) Vessel length (in $\mu m$) vs Strahler order drawn from  Ref.~\cite{nordsletten} (Table 5) and from our generated tree, respectively. (e) and (f) Number of vessels (in log scale) of particular Strahler order  drawn from Ref.~\cite{nordsletten} (Table 2) and from our generated tree, respectively. (g) Total cross section area (in $mm^2$) vs Strahler order from our generated tree. (h) Number of afferent arterioles (AA) branching from the parent vessel of  Strahler order from our generated tree. }
\label{fig_numerical_val}
\end{figure}

Numerical validation is done by comparing the morphometric properties as shown in Fig.~\ref{fig_numerical_val}, such as the vessel radius, branch length, and Strahler order distributions of our reconstructed network with data from a real renal arterial tree collected in a rat kidney \cite{nordsletten}. For example, Ref.~\cite{nordsletten} (Table 2) lists the mean and standard deviation of vessel radius in each Strahler order, as well as the estimated number of vessels in each order. Specifically, radius has been found to increase exponentially with Strahler order both in the real renal arterial tree and in our simulated tree as shown in Fig.~\ref{fig_numerical_val}a and Fig.~\ref{fig_numerical_val}b. This is contrary to \cite{shen2021mathematical} which shows that the result from GCO and from data on the brain vasculature both follow a linear increase of radius with Strahler order. This indicates that although the cost function and general process are similar among organs, GCO is able to adjust based on the distinct geometrical features of each organ. On the other hand, we note that the exact values of the generated radius deviate a bit from the values reported in the literature \cite{nordsletten}, as can be seen for the radius of vessels of Strahler order 10 in the two figures. This is due to Murray's law being only an approximation. In our initialization process, we assume 30K afferent arterioles with $r_0 \sim \mathcal{N}(10.08, 0.14)$ derived from Ref.~\cite{nordsletten} (first row of Table 2). Given strict compliance with Murray's law, the root radius (radius at Strahler order 10) can be computed analytically by $r_{10}=\sqrt[3]{\sum_{i=1}^{n=30000} r_{0, i}^3}$ regardless of branching patterns, which will give a mean value around  $\mu(r_{10}) \approx 313.21\ \mu m$. This number matches our result but deviates from  Ref.~\cite{nordsletten} (last row of Table 2) where $r_{10} \sim \mathcal{N}(216.10, 4.74)$. This may indicate that although our results show that radius should not be optimized in the cost function defined in Eq.~\eqref{eq_final_cost}, Murray's law needs to be adjusted accordingly at some point. In fact,  Ref.~\cite{nordsletten} (Fig. 14) indicates that the renal arterial tree does follow Murray's law in general, but also with noticeable variations.  

Similarly, we plot the vessel length for each Strahler order, both from the literature \cite{nordsletten} and from our result as shown in Fig.~\ref{fig_numerical_val}c and Fig.~\ref{fig_numerical_val}d. Both the data in the literature and our work show that vessel length has a poor correlation with Strahler order. However, they both show a maximum at order 8, indicating that the large arteries usually are longer, but also that they branch fast when being close to the root, resulting in a decrease of length with increasing Strahler order.

Moreover, we plot the number of vessels vs the Strahler order, both from the literature \cite{nordsletten} and from our result as shown in Fig.~\ref{fig_numerical_val}e and Fig.~\ref{fig_numerical_val}f. The data from both the literature and our work show an exponential decrease in the vessel numbers vs the Strahler order. As a result, both the vessel numbers from the literature and from our generated tree fit very well to a straight line in log scale. 

We further plot the total cross-sectional area vs the Strahler order from our GCO output in Fig.~\ref{fig_numerical_val}g, where the corresponding plot of real anatomical data is given in Ref.~\cite{nordsletten} (Fig. 12). 
In agreement with experimental observations, the total cross-sectional area in the generated vascular tree decreases although the mean radius increases exponentially with Strahler order. 


A final interesting property is the Strahler order of the parent vessel of each afferent arteriole, shown in Fig.~\ref{fig_numerical_val}h. Specifically, a parent Strahler order 1 means that afferent arterioles (order 0) branch from terminal arteries (order 1). This case indeed consists of most of the scenarios, but it also demonstrates other possibilities, where afferent arterioles can branch from larger vessels. Specifically, the parent vessels of afferent arterioles have Strahler orders from 1 to 8, meaning that in our model, afferent arterioles can branch from any of the larger vessels, except the largest vessels with Strahler order 9 and 10. This characteristic has been shown to be crucial for the supply of blood to the glomeruli at a sufficient pressure \cite{marsh2019nephron,postnov2016renal}.



We do acknowledge that there are several approximations involved in the whole process, e.g., all the assumptions defined in Section~\ref{para_assumptions}, the approximation algorithm involved in the splitting process in Section~\ref{para_gco_process}, as well as the cortex approximation in Section~\ref{para_image_prior}. However, we believe most of them are inevitable, and we are the first to be able to mathematically model the whole arterial structure in a rat kidney, and to show a clear correspondence with experimental data \cite{nordsletten}, despite all the approximations.

\subsection{Physiological features}
\begin{figure}[htb]
\centering
   \includegraphics[width=.8\textwidth]{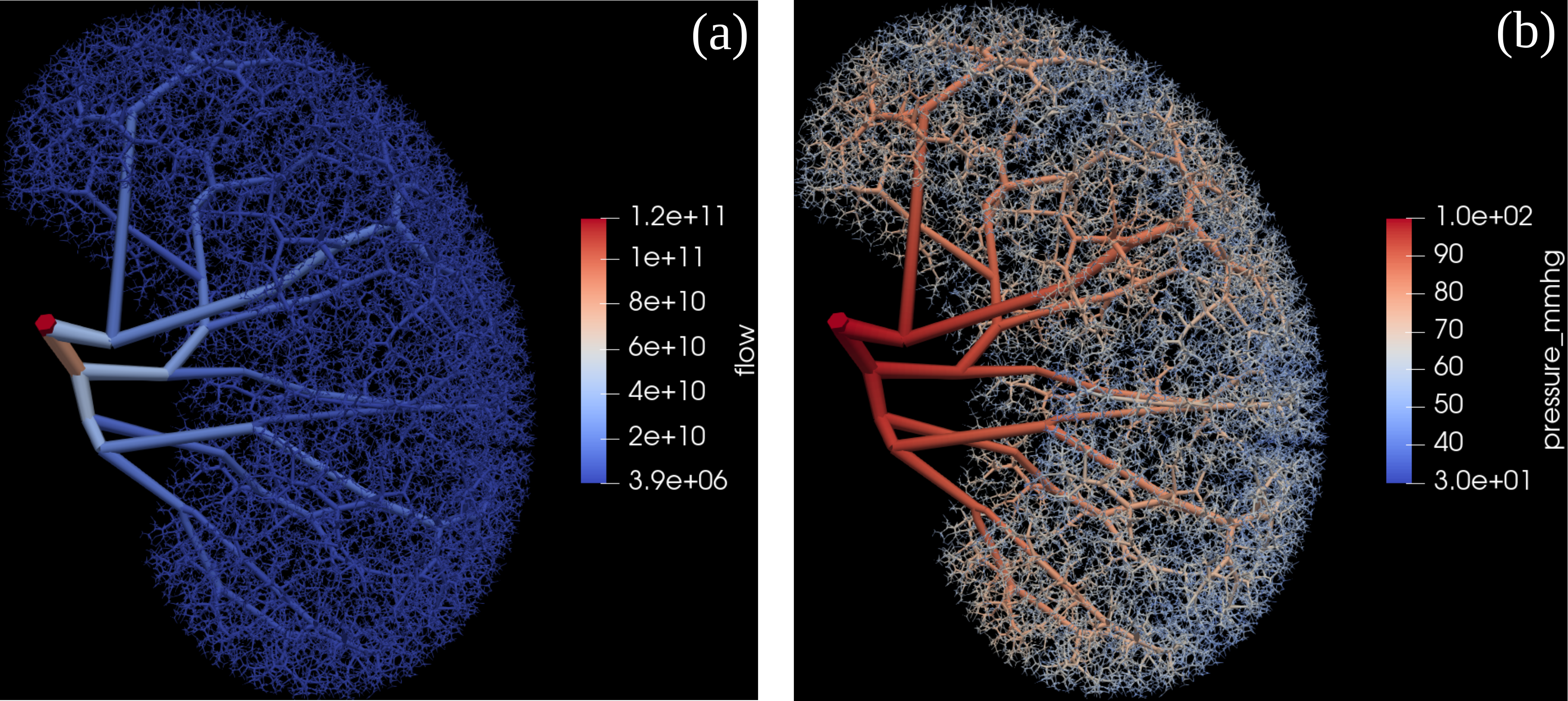}
 \caption{(a) Visualization of blood flow (in $\mu m^3/s$) distributed across the network: from $1.2 \times 10^{11} \mu m^3/s$ (7 $ml/min$) in renal artery to $4 \times 10^{6} \mu m^3/s$ (4 $nl/s$) in afferent arterioles (AA). (b) Visualization of pressure (in mmHg)  distributed across the network, ensuring smooth pressure drop from 100 mmHg at the inlet to a minimum of 30 mmHg at the end of afferent arterioles (AA). Each vessel is visualized by a separate cylinder with a thickness corresponding to its radius but color coded differently by the flow (a) and pressure (b).}
\label{fig_whole_vis_3D}
\end{figure}

\begin{figure}[htb]
\centering
   \includegraphics[width=1.0\textwidth]{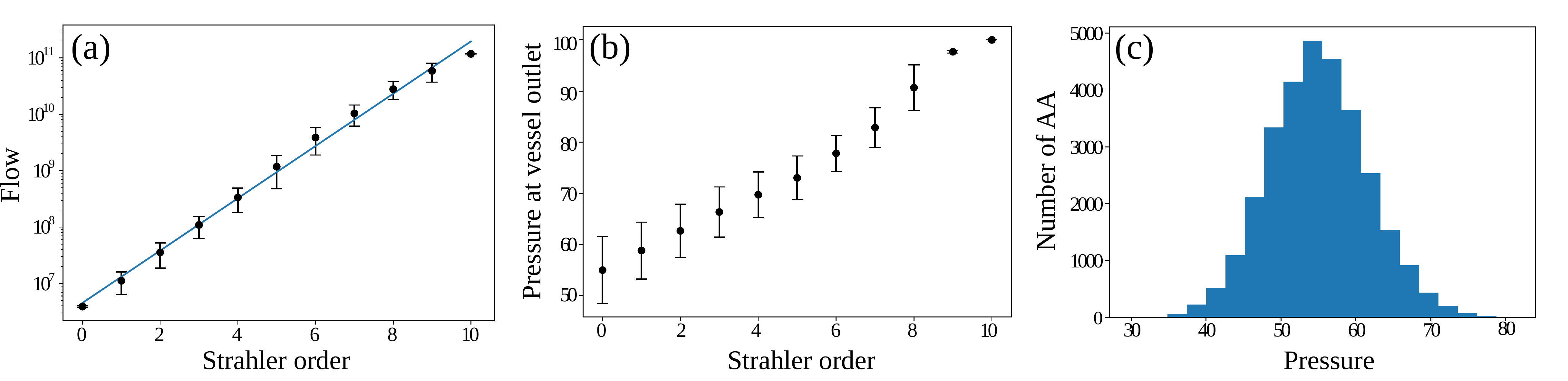}
 \caption{Physiological features of the generated renal vascular arterial network of the whole rat kidney. It shows good agreement with physiological measurements reported in the literature. (a) Blood flow distribution (in $\mu m^3/s$) vs Strahler order
(in log scale). (b) Pressure at the outlet of each vessel (in mmHg) vs Strahler’s order, and (c) Histogram of pressure distribution (in mmHg) among afferent arterioles.}
\label{fig_whole_vis}
\end{figure}

To examine the physiological properties of the generated tree, we plot the blood flow and pressure distribution in Fig.~\ref{fig_whole_vis_3D} and Fig.~\ref{fig_whole_vis}. The flow associated with each vessel is derived from the zero-addition rule (cf. Eq.~\eqref{eq_flow_conserve}) and the assumption of equal flow distribution among the afferent arterioles. As shown in Fig.~\ref{fig_whole_vis_3D}a, the blood flow over our generated renal arterial network ranges from $1.2 \times 10^{11} \mu m^3/s$ (7 $ml/min$) in the renal artery to around $4 \times 10^{6} \mu m^3/s$ (240 $nl/min$) in afferent arterioles (AA).

We further plot the flow in each vessel vs the Strahler order in Fig.~\ref{fig_whole_vis}a (in log scale), which shows a clear exponential increase in flow with Strahler order. Although we have found no literature on such statistics in a real rat kidney, this exponential increase is in close agreement with \cite{kassab2019network} which measures the coronary blood flow vs Strahler order.

The pressure drop $\Delta p_i$ along each vessel $i$ in the generated vascular tree can be computed by Hagen–Poiseuille’s law, cf. Eq.~\eqref{eq_Poiseuille}. Therefore, given the boundary condition of the inlet pressure $p_0$, the exact pressure value at every node along the generated tree can be computed by a simple breadth-first-search with $p_{i+1} = p_i - \Delta p_i$, where $p_{i+1}$ and $p_{i}$ denote the pressure at the outlet and inlet of vessel $i$ respectively. From the literature \cite{casellas1993autoregulation}, pressure in the renal artery is around 90 to 110 mmHg, we hereby assumed an inlet pressure $p_0 = 100$ mmHg.

The pressure at each node in 
Fig.~\ref{fig_whole_vis_3D}b
shows a smooth decrease from $100$ mmHg to a minimum of around $30$ mmHg along the network without abrupt changes, indicating that the reconstructed vascular network produces physiologically feasible hemodynamic behaviors.

We plot node pressure (in mmHg) at the outlet of each vessel vs Strahler order of the generated renal arterial network in Fig.~\ref{fig_whole_vis}b, which shows a smooth and linear increase from around 55 mmHg at the end of afferent arterioles (Strahler order 0) to near 100 mmHg at the end of vessels with Strahler order 9, which then becomes flatter at the last order. This is expected because the root vessel has very small length $l$ and large radius $r$, resulting in a small pressure drop $\Delta p = \frac{8 \mu l Q}{\pi r^4} $ from Poiseuille's equation.

We further plot the histogram of the pressures at the end of the afferent arterioles in 
Fig.~\ref{fig_whole_vis}c. Experimental data \cite{casellas1993autoregulation} shows that the pressure at the end of the afferent arteriole is around 50-55 mmHg, which is in close agreement with the mean value from our result. However, since we only have a reasonable structure but have not modeled the active regulation of pressure at this site, the histogram shows a wider distribution than found experimentally. In the future, such regulation and interaction among contiguous afferent arterioles \cite{marsh2013multinephron,postnov2016renal} need to be modeled to include the fine-tuning of the pressure and radius of afferent arterioles.


\section{Discussion}
We propose a hybrid framework for the reconstruction of the arterial vascular network of a rat kidney. The framework generates a full-scale 3-D vascular tree model based on a modified Global Constructive Optimization algorithm while taking image-based priors from a subject-specific scan. The hybrid method preserves subject-specific information by taking both the kidney's shape and the main artery segmentation from micro-CT images of a real rat into the initialization step. The reconstructed vascular tree structures from the GCO output are validated by showing surprisingly close morphometric agreement with existing anatomical data of a real rat kidney from the literature \cite{nordsletten} and a smooth pressure distribution throughout the vascular tree, which is also in agreement with \cite{casellas1993autoregulation}. Importantly, the reconstructed structure is not a simple bifurcating tree, a structure which previously has been shown to be insufficient for supplying blood to the glomeruli at a sufficient pressure \cite{postnov2016renal}. Instead, in agreement with previous work on the structure of the renal vascular tree \cite{marsh2019nephron}, afferent arterioles arise not only from terminal arteries (Strahler order 1), but also from all arteries of higher order, except for the largest arteries (Stahler order 9 and 10) (cf. Fig.~\ref{fig_numerical_val}h). Therefore, the proposed method can generate both morphometrically correct and physiologically feasible vascular trees while respecting the prior information from the real subject scan. 


Modifications of subprocesses can easily be integrated into our framework. For example,  although the current image prior requires a semi-automated segmentation of the main arteries, one can replace it with state-of-the-art Convolutional Neural Networks, e.g., UNet \cite{ronneberger2015unet} to do auto segmentation if one has accurate training data. Similarly, if renal cortex segmentation is feasible in a different scanning setting, one could skip the cortex extraction (Eq.~\ref{eq_cortex_extraction}) step and directly do the sampling of terminal nodes over the segmented cortex.

Future work will be to apply the reconstructed network in the areas described in the previous section, e.g., to model the active regulation of pressure over our generated arterial tree \cite{marsh2013multinephron,postnov2016renal}. In particular, we assume that our model can generalize to the human kidney as well, given similar micro-CT scans with similar resolution. The renal medulla imposes intrinsic constraints on artery growth, which must be addressed in the vanilla CCO or GCO process. However, since our network starts with a prebuilt tree, the vessels will never pass these regions as long as the vessel segmentation is accurate to a certain extent, giving piece-wise convex regions in the initialization process. 

Another future direction is to create a synthetic renal vessel dataset by generating the ground truth segmentation labels corresponding to the generated tree. To create such an image dataset, we need an inverse task to remap the reconstructed vascular tree back to a smooth surface mesh or binary label map. The detail of such process and an example of image-label pair (Fig. S\ref{fig_flow_chart}) are given in the Supplementary. Hopefully, these artificially generated vessel images can be used to pre-train a deep-learning-based segmentation network for transfer learning or to train a Generative Adversarial Network (GAN) for domain transfer. As an example, Menten et al. \cite{menten2022physiology} recently applied CCO to synthesize retinal vascular plexuses and generated corresponding Optical coherence tomography angiography (OCTA) images by emulating the OCTA acquisition process. They showed that these fake data can successfully pretrain a retinal vessel segmentation network to segment real OCTA retinal images. 

Upon finishing our work, we also notice that each subtree inside each piece-wise convex region after the initialization step of our GCO method (cf. each colored subtree in Fig.~\ref{fig:whole_process}h) is independent of each other. Specifically, the leaf nodes that are connected to a certain node of the pre-built tree in the initialization step will always belong to the successors of that node. Although they all belong to a larger tree with a single root vessel, each subtree can be optimized independently in parallel before being merged together in the end. This parallelization has not been implemented explicitly, which should also be a future direction to speed up the whole computation.

The pair-wise coupling of the arterial and venous systems is not trivial to integrate, when in the future extending the framework to cover both arteries and veins. Currently, it is only possible to independently generate two individual trees. This does not capture the pair-wise coupling of arteries and veins, nor does it avoid the early intersection of the two trees.  Kretowski et al. applied a CCO-based approach to creating complementary hepatic arterial and venous trees \cite{kretowski2003physiologically}. They detect and avoid intersections between the two trees explicitly during the growing process by adjusting the radius of each vessel at the expense of violating Murray's law (Eq.~\eqref{eq_murray}). This process is time-consuming and needs modification over the GCO process, where vessels are optimized on a larger scale.

Finally, we can generate more realistic trees by utilizing more vessels from the segmentation of the scans, or by having a better estimate of the renal cortex region in the initialization step. Currently, we are deliberately removing a large portion of the extracted centerline to only preserve the main arteries, since the other parts are prone to noise and difficult to detect by later auto-segmentation methods. If we have better segmentations, these small segments could also be used to guide the reconstruction of subject-specific vascular networks. On the other hand, one could also experiment with how small a portion of a pre-built tree we need in the initialization step of GCO before it will violate anatomical constraints, e.g., pass through the renal pyramid or outside the kidney structure without the piece-wise convex premise.
Another potential future direction involves integrating large vessels in a totally different manner. Instead of a pre-built tree, the segmentation could also be incorporated in the cost function as gravitation to guide the whole process by ``pulling'' the intermediate nodes close to their positions. However, it is non-trivial to add such pull-force to Eq.~\eqref{eq_final_cost} while balancing the other constraints. It will probably involve lots of hyper-parameter tuning to find such a balance.

\section{Methods}
The vanilla CCO algorithm works by iteratively adding a new edge (vascular segment). After each addition, the newly created bifurcation is locally remodeled and all tree radii are adjusted geometrically. Such remodeling is clearly inefficient when the vessel structure becomes huge like in the kidney. In contrast, we adopt the alternative GCO algorithm as our backbone model. This method overcomes the problems of the CCO by starting with a fully connected tree, where the leaves are usually either defined on a regular grid or randomly positioned within the organ hull \cite{georg2010global}. In our case, the leaves are sampled from the estimated renal cortex (Fig.~\ref{fig:whole_process}c) as detailed in Section~\ref{para_leaf_node}. It performs a multi-scale optimization to find an optimal tree for all leaf nodes simultaneously and introduces a global pruning operation after each iteration to produce a new tree with better global branching structures.

\subsection{Assumptions and objectives}
\label{para_assumptions}
Several assumptions have to be made for the whole process of reconstructing vascular trees. The first assumption forms the basis of the mathematical modeling of any vascular tree. The second assumption is the rationale for a Poisson disk sampling of terminal nodes (glomerulus), which will be discussed in Section~\ref{para_leaf_node}. The other assumptions are necessary to satisfy Murray's law in Eq.~\eqref{eq_murray} and Poiseuille's equation in Eq.~\eqref{eq_Poiseuille}, which are the basis in the derivation of the cost function Eq.~\eqref{eq_final_cost}, as well as in the merging and splitting process.
\begin{itemize}
    \item A vascular tree is modeled as a collection of connected, straight cylindrical tubes, indicating constant radius and no curvature before branching.
    \item All the renal arteries end in the renal cortex with a certain perfusion territory.
    \item Blood is incompressible and Newtonian, and blood flow is laminar. 
    \item Pressure drop due to branching is negligible.
    \item Flows are equally distributed among each terminal vessel.
\end{itemize}
Given such assumptions, a vascular tree is modeled by a directed acyclic graph $\mathcal{G} \equiv (\mathcal{V}, \mathcal{E})$ where $\mathcal{V}$ is a set of nodes in the two endpoints of each vessel centerline with node features being its coordinates in Euclidean space, and $\mathcal{E}$ is a set of directed edges which form connected, tree structure, representing each cylinder with its radius and flow as the edge feature. Note that length is not modeled as an edge feature but rather derived from the Euclidean distance between the two end nodes of each edge. The goal is to find a tree that minimizes the system's overall cost function while fulfilling the constraints. Specifically, given the position of a single root node $s$, and $n$ leaf nodes $l_i$, 
the goal is to find a tree $\mathcal{G} \equiv (\mathcal{V}, \mathcal{E})$ that contains $s$ and $l_i \in \mathcal{V}$ with minimum cost defined in the next subsection and fulfill constraints by introducing new intermediate nodes $v_i \notin \{s, l_1, \cdots, l_n\}$ and connections (edges) $\mathcal{E}$.

In our work, we propose a novel way to integrate image priors into the initialization of GCO, so that the input is no longer a single root with $n$ leaves, but a pre-built tree $\mathcal{G}_0 \equiv (\mathcal{V}_0,\mathcal{E}_0)$ with $s \in \mathcal{V}_0$ that already covers the main arteries.

\subsection{Physiologically based cost functions}
\label{para_cost_function}
Our reconstruction method generates subject-specific arterial vascular networks $\mathcal{G} \equiv (\mathcal{V}, \mathcal{E})$ under the optimality assumption that the network structures will maintain adequate blood perfusion with minimal total expense along all its edges, which can be approximated by the total sum of the local cost at each branching node $v$:

\begin{equation}
    C(\mathcal{G})  \equiv \sum_{v} C_{\text {local}}(v) 
    \equiv 
    \sum_{v}  \sum_{e \in \mathcal{B}_{v}} C(e)
    \,
\label{eq_total_cost}
\end{equation}
where $\mathcal{B}_{v}$ denotes the set of all the incident edges of node $v$ 

\begin{figure}[htb]
\centering
    \includegraphics[width=.6\linewidth]{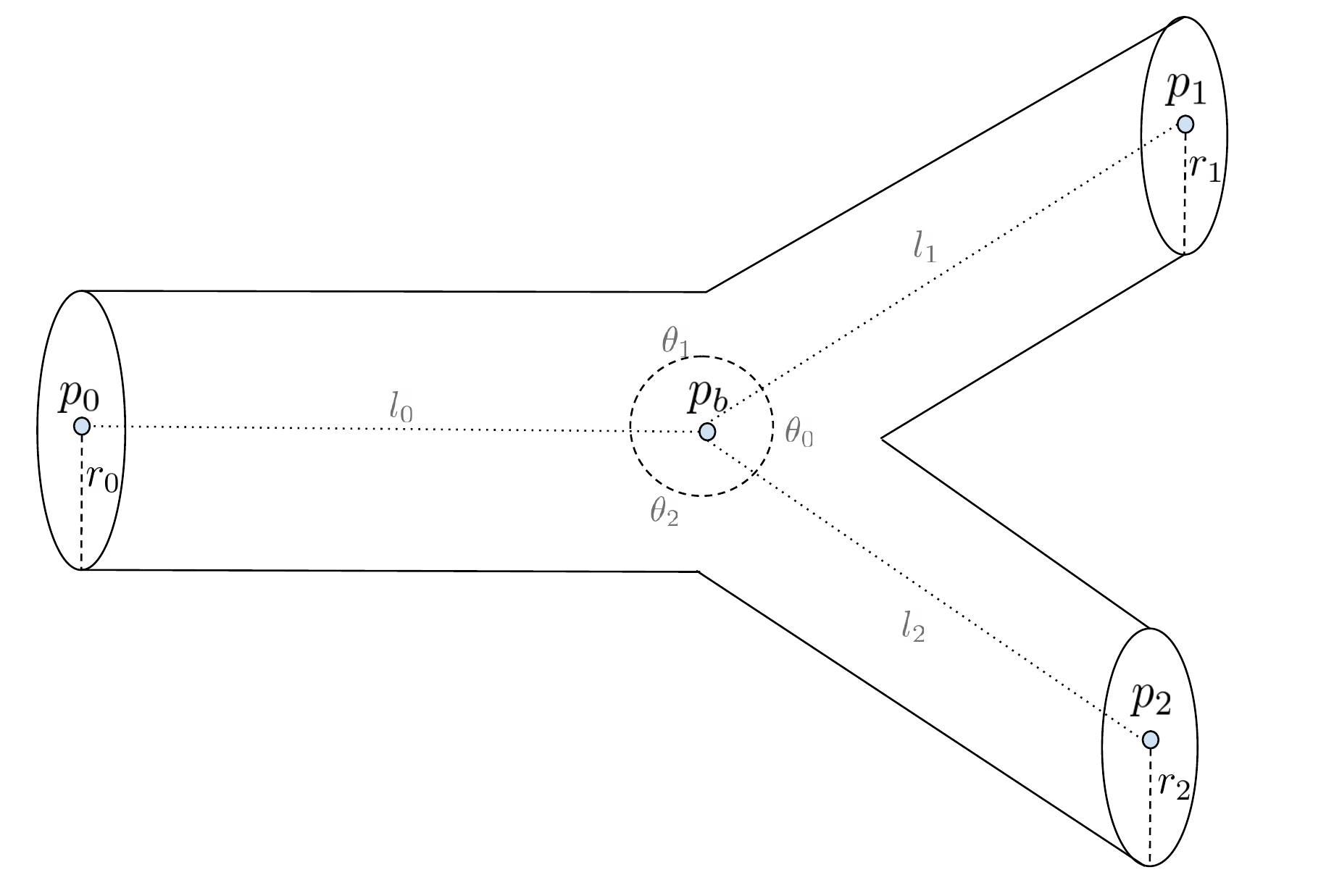}
  \caption{A typical vessel branching model in bifurcation, which is uniquely defined by the locations of three end nodes ($p_0, p_1, p_2$), the location of the bifurcation node ($p_b$), and the radii of the three incident edges ($r_0, r_1, r_2$). Length ($l_s$) and branching angles ($\theta_s$) are not modeled explicitly but can be trivially derived.}
\label{fig_bif_illu}
\end{figure}

A typical branching model is shown in Fig~\ref{fig_bif_illu}. Note that the GCO model does not enforce bifurcation explicitly and can indeed model any number of branches $>2$. Still, branches other than bifurcation or trifurcation are rarely seen in the final result, because they usually incur a higher cost. In each branching model, an optimal branching point is positioned w.r.t. fixed neighboring edge radii and neighboring node positions to minimize the cost function. Following the work of Tekin and Shen et al. \cite{shen2021mathematical,Elif_tekin2016vascular}, we incorporate both the material cost ($M_{loss}$) and power cost ($P_{loss}$), resembling the biological infrastructure cost to build the vessel and the power dissipated during blood circulation, respectively. Therefore, $C_{\text {local}}(v)$ is a weighted combination of the two costs:

\begin{equation}
\label{local_m_p}
        C_{\text {local}}(v) \equiv
    w_c \, M_{loss}(v) + w_p \, P_{loss}(v)
\end{equation}
where $M_{loss}(v)$, which expresses the amount of materials that constitute the blood in the vessels, is only dependent on the intravascular volume of the arterial tree. It is given by 

\begin{equation}
\begin{aligned}
    M_{loss}(v) \equiv \sum_{e \in \mathcal{B}_{v}} M(e) \equiv \sum_{e \in \mathcal{B}_{v}} \pi r_e^2 l_e 
    \,
\end{aligned}
\label{eq_material_cost}
\end{equation}


Analog to electric power, the power dissipated during blood circulation ($P_{loss}(v)$) is defined by the product of flow $Q_{e}$ (analog to electric current) and pressure drop $\Delta p_e $ (analog to potential difference),  

\begin{equation}
\label{eq_ploss_orig}
    P_{loss}(v) \equiv \sum_{e \in \mathcal{B}_{v}} Q_{e} \Delta p_e 
\end{equation}
where Hagen–Poiseuille's law gives the pressure drop $\Delta p_e$ necessary to overcome the resistance to flow due to the viscosity, $\mu$,  of the blood in an individual blood vessel,

\begin{equation}
    \Delta p_e 
    = 
    \frac{8 \mu l_e Q_e}{\pi r_e^4} 
    \label{eq_Poiseuille}
    \,
\end{equation}

Combining Eq.~\eqref{eq_ploss_orig} and~\eqref{eq_Poiseuille} gives



\begin{equation}
\begin{aligned}
    P_{loss}(v) \equiv \sum_{e \in \mathcal{B}_{v}} Q_e^2 \frac{8 \mu l_{e}}{\pi r_{e}^{4}}
    \,
\end{aligned}
\label{eq_power_cost}
\end{equation}

The total cost at each branching $v$ is a weighted combination of material cost Eq.~\eqref{eq_material_cost} and power cost Eq.~\eqref{eq_power_cost},

\begin{equation}
\begin{aligned}
    C_{\text {local}}(v)   
    &\equiv
    w_c \, M_{loss}(v) + w_p \, P_{loss}(v) 
    \,
    \\
    &= 
    w_c \, 
    \left(
    \sum_{e \in \mathcal{B}_{vx}}
        \pi r_e^2 l_e 
    \right)
    + 
    w_p \, 
    \left(
      \sum_{e \in \mathcal{B}_{v}} Q_e^2 \frac{8 \mu l_{e}}{\pi r_{e}^{4}}
    \right)
    \,
    \\
     &= 
    \sum_{e \in \mathcal{B}_{v}} \left(
     w_c \pi r_e^2 l_e 
         + w_p Q_e^2 \frac{8 \mu l_{e}}{\pi r_{e}^{4}}
    \right)
    \,.
\end{aligned}
\label{eq_final_cost}
\end{equation}

The current flow follows the simple zero-addition rule from Kirchhoff's first law, from which we derive the relation,

\begin{equation}
    Q_p = \sum_{e \in \mathcal{B}_{v}/p} Q_e
    \label{eq_flow_conserve}
\end{equation}
where $p$ denotes the parent edge in $\mathcal{B}_{v}$ of node $v$.
Note that radii are not optimized but completely follow Murray's law,

\begin{equation}
    r_p^3 = \sum_{e \in \mathcal{B}_{x}/p} r_e^3
    \,
    \label{eq_murray}
\end{equation}

Although we do understand that Murray's law is only an approximation, we found that optimizing radii, e.g., by integrating an equality constraint into the optimization process as proposed by \cite{shen2021mathematical} only deteriorates the result.

In our experiment, we set the weight factors $w_c=5 \times 10^{4} \mathrm{~J} \mathrm{~s}^{-1} \mathrm{~m}^{-3} = 5 \times 10^{-8} N \, {\mu m^{-2}} s^{-1}$, 
and $w_p=1$, as we have found that they result in the same magnitude of the two cost terms. We also adopt constant viscosity $\mu=3.6 \times 10^{-3} \mathrm{~Pa} \, \mathrm{~s} = 3.6 \times 10^{-15} \mathrm{N \, s  \, \mu m^{-2}}$, and inlet flow $Q_0 = 7ml/min = 1.167 \times 10^{11} \mu m^3 \, s^{-1}$ with values from the literature \cite{stannov2019interactions,ronn2017glucagon}. Assuming equal flow distribution over terminal vessels, flow at afferent arterioles can be calculated as $Q_t = \frac{Q_0}{N}$ where $N$ is the number of afferent arterioles. In our case, $Q_t =\frac{1.167 \times 10^{11} \mu m^3 \, s^{-1}}{3 \times 10^4}= 3.89 \times 10^6 \mu m^3 \, s^{-1} = 3.89 \ nl \, s^{-1}$, complying with the literature where $Q_t \approx \ 4 nl \ s^{-1}$ \cite{marsh2005frequency}. Note that the physics units and voxel size need to be consistent with each other to ensure that $M_{loss}$ and $P_{loss}$ are on the same scale. The above units give both $M_{loss}$ and $P_{loss}$ in the scale of $N \, \mu m \, s^{-1}$ ($\mu W$).


\subsection{Global Constructive Optimization algorithm}
\label{para_gco_process}
The Global Constructive Optimization (GCO) algorithm includes the following steps, which are iterated multiple times before convergence except for the first initialization step. Please refer to \cite{georg2010global} for a more detailed explanation of the process.
\begin{enumerate}
    \item 
\textit{Initialization.} In the original GCO initialization, each sampled terminal node is connected to a single user-defined root node, thus completely ignoring subject-specific information. In our hybrid framework, the sampled terminal nodes are connected to the main arteries derived from the patient's scan. Details of the sampling process and main arteries retrieval are explained in the next subsection (\ref{para_image_prior}).
\item
\textit{Relaxation.} The relaxation process finds the best location for each branching node through optimization by minimizing the overall cost function defined in Section~\ref{para_cost_function}.
\item
\textit{Merging.} Merging involves contracting the edge between two neighboring nodes. It is applied when the ratio between the shortest incident edge of a node and the second incident edge is within a threshold, which usually happens when relaxation places a node at the same location as one of its neighboring nodes.
\item
\textit{Splitting.} Splitting is done whenever creating a new intermediate node and reconnecting a subset of the original child neighbors $S \in \mathcal{B}_v/{p}$ introduces a lower cost. Usually, this condition is fulfilled at a node with too many edges, indicating that bifurcation is implicitly imposed on the modeling. However, finding the optimal subset is in $\mathcal{O}(n!)$ thus NP-hard. Instead, an approximation algorithm is applied by first finding a subset $S_1 \in S $ with two edges that introduce the lowest cost. A new edge is then iteratively added to $S_n$ from $S-S_{n-1}$ if it introduces a lower cost. This approximation has a complexity of $\mathcal{O}(n^2)$ thus much more efficient. It is worth mentioning that this operation is still more computationally heavy than the actual optimization step in relaxation.
\item
\textit{Pruning.} A pruned tree $\mathcal{G}_l \equiv (\mathcal{V}_l, \mathcal{E}_l)$ is created from $\mathcal{G}$ by removing all edges with an order smaller than some threshold. This process will only keep the large branches generated from each iteration and produce a new tree with a better global branching structure in the next iteration. All the leaf nodes that are removed in this operation are reconnected to the nearest node in the pruned tree. The modification we make here is that each leaf node can only be reconnected to the subtree that it belongs to in the initialization step.

\end{enumerate}

\subsection{The Image Priors}
\label{para_image_prior}
In general, our proposed hybrid way of utilizing image-based priors involves two segmentation maps from the kidney scans, to begin with, as shown in Fig.~\ref{fig:whole_process}: segmentation of large arteries $Y_a$ (Fig.~\ref{fig:whole_process}e), and segmentation of the whole kidney structure $Y_w$ (Fig.~\ref{fig:whole_process}a). 
The segmentation of large arteries is obtained using a semi-automated approach \cite{andersen_charlotte_2021evaluation}. Whole kidney structure segmentation, however, is obtained by simple thresholding since the ex-vivo micro-CT scan makes it extremely easy to separate the kidney from the background. These two segmentation maps are used in the following two tasks respectively for the initialization of the whole GCO process. Fig.~S\ref{fig_flow_chart} in Supplementary shows the flowchart of the process.

\subsubsection{The Leaf Node Sampling}
\label{para_leaf_node}
The segmentation of the whole kidney structure $Y_w$ is used to sample terminal nodes (glomerulus) where the arteries end (Fig.~\ref{fig:whole_process}d). Since the arteries end in the cortex region rather than only on the surface, to mimic the anatomical rules, several more steps are necessary. 

\bmhead{Cortex Approximation via Erosion} In theory, renal arteries end in the renal cortex, which requires a cortex segmentation to sample from. Since the cortex is not visible from our micro-CT scan, it is approximated by a certain distance ($R_1 \approx 2 \ m m$) away from the surface by assuming equal thickness. This process can be easily obtained by the subtraction of a mathematical erosion ($\bullet$) applied to $Y_w$, as shown in the yellow regions in Fig.~\ref{fig:whole_process}b.

\bmhead{Inner Region Removal} To avoid sampling terminal artery nodes around the renal artery, all the regions near a certain distance to the root node $v_r$ are removed. This is  accomplished by imposing a ball centered at $v_r$ ($R_2 \approx 5.65 \ mm $), as shown in the green regions in Fig.~\ref{fig:whole_process}b. In summary, cortex segmentation is a set of points 

\begin{equation}
     Y_c 
     \equiv 
     \left\{
        ~\mathbf{x}~
        \mid
        ~
        \mathbf{x} \in Y_a - (Y_a \bullet R_1)~\land~ \|\mathbf{x} - \mathbf{v}_r\|_2 > R_2 
     \right\}
\label{eq_cortex_extraction}
\end{equation}
of which the surface mesh is visualized in Fig.~\ref{fig:whole_process}c.


\bmhead{The Poisson disk sampling} Vessels in any organ follow an anatomical structure that the leaf nodes should cover the entire perfusion territory while avoiding being too close to each other to prevent competition or overlap between vessel branches \cite{keelan2016simulated}. Poisson disk sampling \cite{bridson2007fast_poisson_sample} maintains a minimum distance between sampled points by rejecting points that are too close to each other, which resembles such an anatomical rule very much. In the present model, the minimum distance value can be approximated from the cortex volume and number of points we would like to sample from. We follow the work of Nordesletten \cite{nordsletten} by sampling 30K terminal vessels (number of arteries with Strahler order 0), which results in a minimum  distance of around 270 $\mu m$. Monte Carlo sampling is also adopted to do Poisson disk sampling over the whole cubic volume before filtering out the points from non-cortex regions.

\subsubsection{Large Artery Extraction}
To integrate the large artery segmentation $Y_a$ into the GCO process, vessels need to be modeled by a graph with nodes along the centerline and edges with connectivity information. Therefore, we start by extracting the centerline from $Y_a$ using the Skeletonization method proposed by Bærentzen et al. \cite{baerentzen2021skeletonization_local_separator}. Instead of a binary image with width $1$ at each local foreground voxel, the algorithm outputs a graph data structure $C(Y_a) = \mathcal{G}(\mathcal{V}, \mathcal{E})$ suitable for our needs. However, the extracted centerline (Fig.~\ref{fig:whole_process}f) is an undirected graph and may contain loops, which cannot be directly used. Several preprocessing operations are necessary before initialization.

\bmhead{Minimum Spanning Tree} This first operation removes potential loops by creating a subset of the edges with the minimum total edge weight from  $\mathcal{G}(\mathcal{V}, \mathcal{E})$ that connects all the vertices without any cycles. Here the weight of each edge is the negative of its radius derived from a Euclidean distance transform over the derived centerline to the segmentation. This operation will remove the thinnest edge to break any loop. After this step. The acyclic graph can be converted to a tree (directed acyclic graph) with a simple depth-first-search.

\bmhead{Intermediate nodes removal} As stated in the assumption from Section~\ref{para_assumptions} that each vessel is modeled by a straight cylindrical tube, any intermediate nodes along each single vessel will have to be removed. This will, of course, introduce artifacts to the length computation, but is assumed to be negligible within a reasonable curvature.

\bmhead{Degree Pruning} For each node with more than 4 branches, we only keep a maximum of 4 longest paths, since branching into more than 4 children is not realistic.

\bmhead{Depth Pruning} For each node, we compute its cumulative distance to the root along the tree and only keep nodes up to a certain distance. The rationale is that even though some thin vessels far away from the root are visible from the current segmentation, only large vessel segmentation is trustworthy. Especially if we would like to further adapt deep learning for automatic vessel segmentation, we cannot assume the model to be able to detail the thin vessels.  


\bmhead{Connected Component Decomposition} The two pruning operations may introduce smaller disconnected trees, we thus only keep the largest tree.

\subsubsection{Final GCO Initialization}
For the initialization of GCO (Fig.~\ref{fig:whole_process}h), all the sampled terminal nodes (Fig.~\ref{fig:whole_process}d) are connected to the nearest ending node along the extracted and pre-processed large artery centerline graph $\mathcal{G'}(\mathcal{V}, \mathcal{E})$ (Fig.~\ref{fig:whole_process}g). The radii associated with the terminal vessels are sampled from a Gaussian distribution $r_0 \sim \mathcal{N}(10.08, 0.14)$ derived from literature \cite{nordsletten}, while radii of other vessels are derived from the radii of terminal vessels by Murray's law (Eq.~\eqref{eq_murray}). 

Besides retaining subject-specific information from image priors, the connection to the pre-built tree also makes the complex structure piece-wise convex, making the later constructive algorithms applicable here. Specifically, the connection between any terminal node to the pre-built tree should not enter or cross the renal pyramid, which is hard to satisfy when the pre-built tree is only a single root node. 

\subsection{Ex vivo micro-CT imaging dataset}
The kidney cast was prepared as described in \cite{andersen_charlotte_2021evaluation} in agreement with approved protocols (approval granted from the
Danish Animal Experiments Inspectorate under the Ministry of Environment and Food, Denmark). The rat kidney was \textit {ex vivo} scanned in a ZEISS XRadia 410 Versa $\mu$CT scanner (Carl Zeiss Microscopy GmbH, Jena, Germany) at the following settings: isotropic voxel size 22.6 $\mu$m, 50 kV tube voltage, 0.2 mA current, appertaining LE3 filter, 360$^{\circ}$ scan around the vertical axis with 3201 different projections
(0.112$^{\circ}$ rotation steps) \cite{andersen_charlotte_2021evaluation}. The raw scan has a dimension of $1000 \times 1024 \times 1014$ voxels. To ease the computational overhead, the scan is auto-cropped to $955 \times 508 \times 626$ by an intersected bounding cube of the largest component from simple Otsu's thresholding over maximum intensity projections to three dimensions.

\subsection{Implementation details}

Shen et al. \cite{shen2021mathematical} and Keelan et al. \cite{keelan2016simulated} proposed to optimize the cost function using Simulated Annealing, which is a metaheuristic to approximate the global optimum of a given function. Because of its non-gradient-based nature, this method is usually preferable for problems where gradients are hard to compute. However, we note that the cost function defined in Eq.~\eqref{eq_final_cost} is quite differentiable, meaning its gradient can be easily computed analytically, giving

\begin{equation}
\begin{aligned}
    \nabla C_{\text {local}}(v) 
     &= 
    \nabla 
    \left(
    \sum_{e \in \mathcal{B}_{v}} \left(
     w_c
        \pi r_e^2 l_e 
        + w_p Q_e^2 \frac{8 \mu l_{e}}{\pi r_{e}^{4}}
    \right)
    \right)
    \,
     \\
          &=  \sum_{e \in \mathcal{B}_{v}}    \left(
     w_c \pi r_e^2
         + w_p Q_e^2 \frac{8 \mu }{\pi r_{e}^{4}}
    \right) \nabla \, \|\mathbf{v} - \mathbf{n}_{v, e}\|_2
    \,
    \\
     &= \sum_{e \in \mathcal{B}_{v}}    
     \left(
        w_c \pi r_e^2
        +
        w_p Q_e^2 
        \frac{8 \mu }{\pi r_{e}^{4}}
    \right)
    \frac{
        \mathbf{v} - \mathbf{n}_{v, e}
    }{
        \|\mathbf{v} - \mathbf{n}_{v, e}\|_2
    }
\end{aligned}
\label{grad_final_cost}
\end{equation}
where $\mathbf{n}_{v, e}$ denotes the position in $\mathcal{R}^3$ of the neighboring node of $v$ along edge $e$. Therefore, we apply the standard Broyden–Fletcher–Goldfarb–Shanno (BFGS) method which proves to perform as well and is much faster. Fig.~S\ref{fig_convergence} in Supplementary shows the convergence plot of the GCO process.

 All the backbones are pure NumPy and SciPy-based computation, with graph representation using NetworkX \cite{hagberg2020networkx}. Currently, there is no GPU acceleration. In fact, the optimization process in relaxation is not the bottleneck. As discussed in Section~\ref{para_gco_process}, splitting is usually the computational bottleneck and dominates the time complexity, especially in the first few iterations where there are a small number of intermediate nodes each with a large number of neighbors. Moreover, since each branching has to be optimized individually and consecutively, switching to PyTorch with GPU acceleration will not help. The whole pipeline takes approximately 10 hours to reach convergence.

\bibliography{sn-bibliography}

\newpage

\vspace*{15pt}
\hspace*{\fill} {\Large Supplementary Material} \hspace*{\fill}
\vspace{15pt}


\section{Visual animations}
Visualizations of our generated renal arterial tree in 3D gif animations are available at \url{https://github.com/KidneyAnonymous/RenalArterialTree}. They are created by applying a Tube filter in ParaView \cite{ahrens2005paraview} where each tube is assigned with the radius of the generated vessel segment. Different gifs represent the tubes colored by different properties, e.g., radius, flow, and pressure. 

\section{Flow chart}
\begin{figure}[htb]
 \centering
\begin{minipage}[b]{.9\linewidth}
  \centering
  \centerline{  \includegraphics[width=1.0\linewidth]{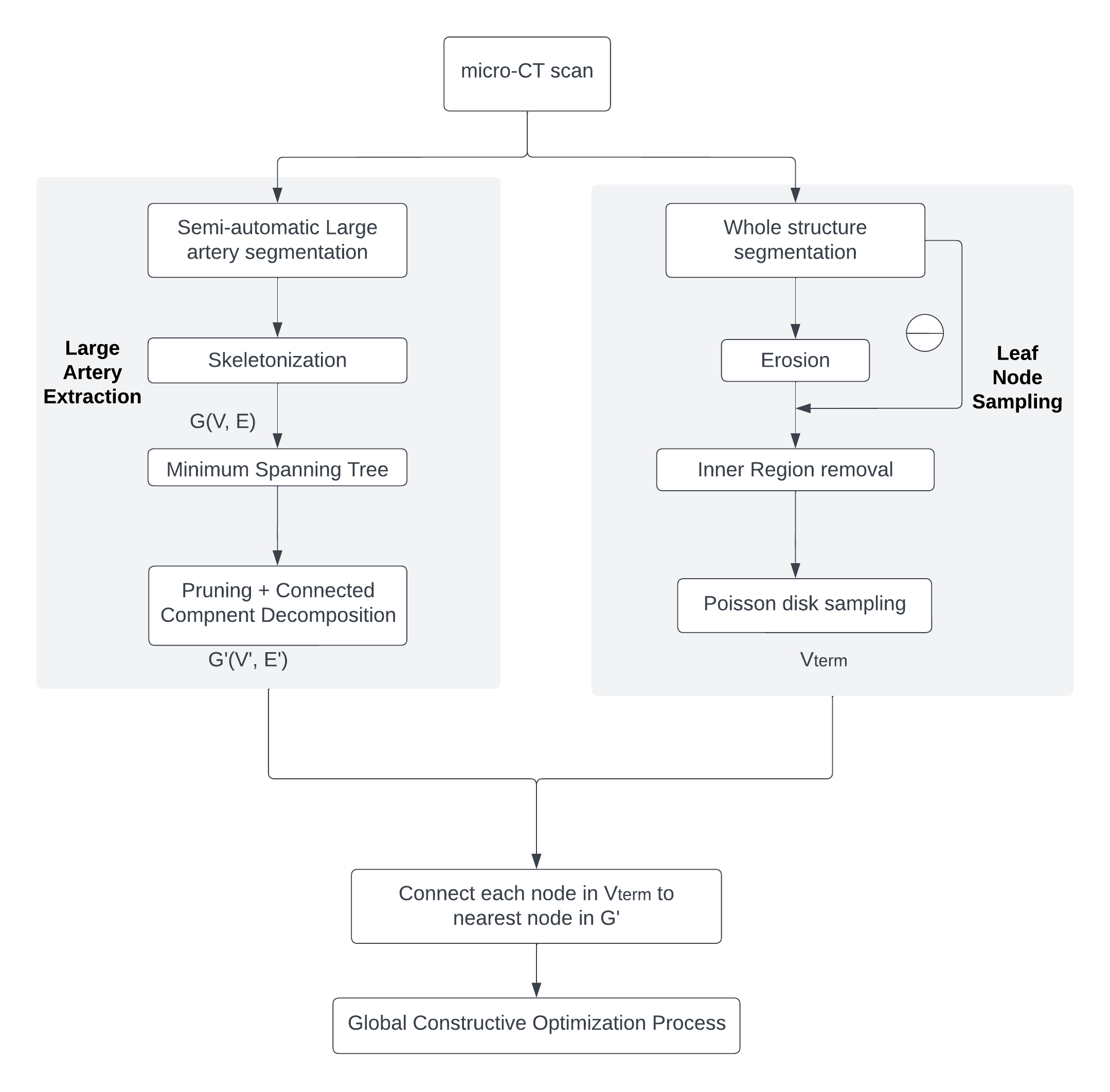}
  }
\end{minipage}
\caption{Flowchart of the complete computational framework where $	\ominus$ denotes element-wise subtraction. In the main text, the whole process is detailed in Section~\ref{para_image_prior}, while the final Global Constructive Optimization process is detailed in Section~\ref{para_image_prior}.}
\label{fig_flow_chart}
\end{figure}

We show the flowchart of how the image priors are integrated into the GCO process in Fig.~\ref{fig_flow_chart}. The micro-CT scan is used to extract a prebuilt large arterial tree and sample leaf nodes from the two segmentation maps respectively with some intermediate steps.
These two outputs are then the inputs to the GCO initialization step to guide the reconstruction of the full-scale arterial tree.

\section{Artificial kidney vessel dataset}
The generated vascular tree from our pipeline will vary between runs since there are several sources of randomnesses involved, e.g., the Poisson disk sampling of terminal vessels, its radii sampled from Gaussian distribution, and the approximation algorithm in the splitting process. We can thus create a synthetic kidney vessel dataset by generating the ground truth segmentation labels corresponding to each generated tree. 
To create such an image dataset, we need an inverse task to remap the reconstructed vascular tree $\mathcal{G} = (\mathcal{V},\mathcal{E})$ back to a smooth surface mesh or binary label map. This is usually much more complicated than simply stacking each individual cylinder together. A reconstructed vascular tree $\mathcal{G} = (\mathcal{V},\mathcal{E})$  hosts the centerline location as well as the radius of its maximal inscribed spheres. The way we generate a smooth surface structure is by constructing a tube function based on the generated tree. The tube function $T: R^3 \rightarrow R$ in Eq.~\eqref{eq_tube_func} is a signed distance transformation that maps every voxel point $\mathbf{x}$ to the tube surface along the centerline \cite{antiga2004robust_bifurcating_tube} such that points inside the vessel have negative values, and positive outside the vessel with smooth transitions. This means a marching cube method can easily generate the surface mesh, and the binary label map can be easily generated by using a $0$ threshold value,

\begin{equation}
    T_{\mathbf{c}, r}(\mathbf{x})
    \equiv 
    \min _{i} 
    \left\{
        \min _{\tau \in[0, L_i]}
        \left\{
            \left\|
            \mathbf{x}-\mathbf{c_i}(\tau)
            \right\|^2-r_i^2(\tau)
        \right\}
    \right\}
\label{eq_tube_func}
\end{equation}
where $\mathbf{c_i}$ denotes the centerline graph of the $i^\text{th}$ line segment, $L_i$ and $r_i$ denote the length and radius of each vessel.
$\tau$ denotes the exact arc length along the line segment starting at node position $\mathbf{s}_i$ and ending at node position $\mathbf{s}_{i+1}$.
%
Hence, 
$\mathbf{c}_i(\tau) \equiv \mathbf{s}_{i+1} \, \left(\tfrac{\tau}{L_i}\right) + \mathbf{s}_{i} \,
\left(1-\tfrac{\tau}{L_i}\right)
$
Assuming both the spatial coordinates and radii are linearly interpolated, $\tau$ can be found analytically by the expression for the closest point to a line in Euclidean geometry \cite{antiga2004robust_bifurcating_tube},

\begin{equation}
    \tau
    \equiv
    \frac{
            \left(
                \mathbf{x}-\mathbf{c}_i\left(s_i\right)
            \right) 
            \cdot
            \left(
                \mathbf{c_i}\left(s_{i+1}\right)-\mathbf{c_i}\left(s_i\right)
            \right)
        }{
            \left\|
                \mathbf{c_i}\left(s_{i+1}\right)-\mathbf{c_i}\left(s_i\right)
            \right\|
        }
    \,
\end{equation}
The process is extremely time-consuming because the tube function has to be evaluated over a large grid on each line segment (vessel) $[\mathbf{s}_i, \mathbf{s}_{i+1}]$ such that the tube function is minimized. A crucial acceleration we make here is that, for each line segment $[\mathbf{s}_i, \mathbf{s}_{i+1}]$, we only compute the tube function inside a bounding cube around it. Specifically, only voxel points within the bounding cube of $[\min(\mathbf{s}_i, \mathbf{s}_{i+1}) - \max(r(\mathbf{s}_i), r(\mathbf{s}_{i+1})), \max(\mathbf{s}_i, \mathbf{s}_{i+1}) + \max(r(\mathbf{s}_i), r(\mathbf{s}_{i+1}))]$ is evaluated for each line segment, where $\min(\mathbf{s}_i, \mathbf{s}_{i+1})$ and $\max(\mathbf{s}_i, \mathbf{s}_{i+1})$ are both in $\mathcal{R}^3$ representing the minimum and maximum coordinates of the two endpoints $s_i$ and $s_{i+1}$ in the three dimensions separately. This simple modification reduces the computational time from weeks to only one minute on a grid size of around $955 \times 508 \times 626$ over a generated vascular tree with around $50$K vessel segments.

\begin{figure}[htb]{}
  \centering
\includegraphics[angle=0, trim={0 0 0 0}, clip, width=\linewidth]{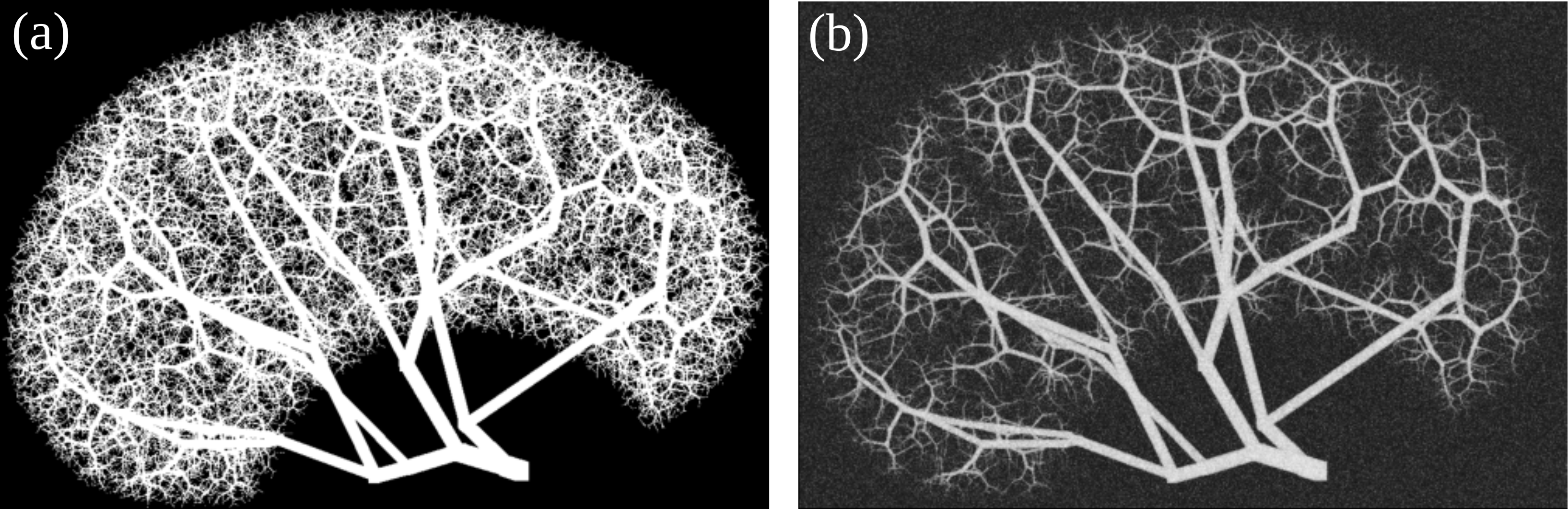}

\caption{Example of the maximum intensity projection of a generated vessel label map from the reconstructed tree (a) and the synthesized image with random Gaussian and Salt\&Peper noise (b). Note that they are mapped back to the original voxel space of $22.6 \ \mu m$, meaning that some small vessels are not visible.
}
\label{fig_fake_image}
\end{figure}

The corresponding scan image can be roughly created by adding some noise to the label maps based on micro-CT characteristics. Fig.~\ref{fig_fake_image} shows the example of a generated label map from the reconstructed tree by applying a threshold on the tub function as well as the synthesized image with some random Gaussian and Salt\&Pepper noise. Pretraining a neural network for renal vessel segmentation would require more information on the scanning details of the micro CT devices to emulate the imaging process and generate realistic noises, which is far beyond the scope of this work.

 \subsection{Convergence plot}
\begin{figure}[H]
 \centering
  \centerline{ \includegraphics[width=0.7\linewidth]{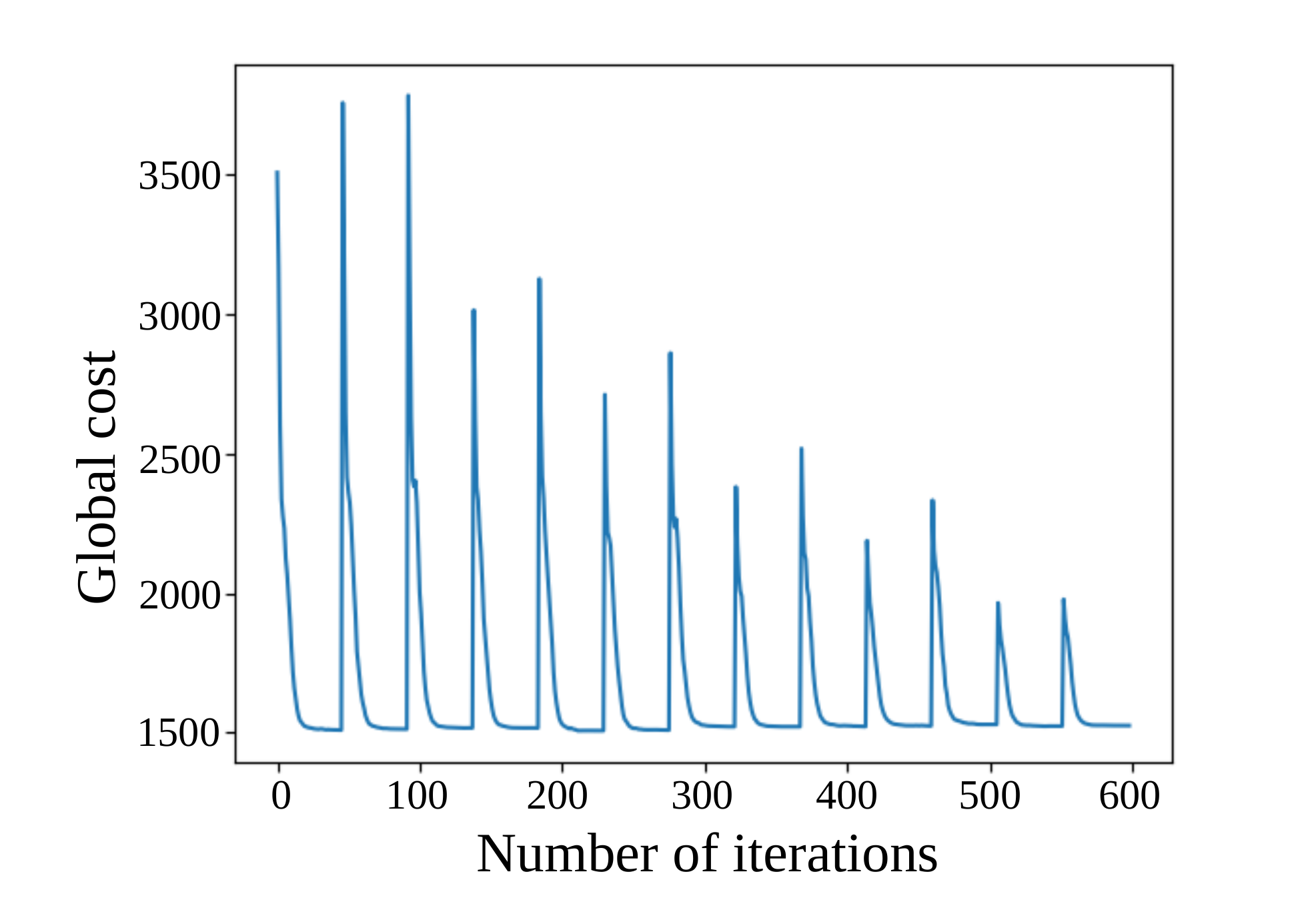}
  }
\caption{Convergence plot over the GCO process.}
\label{fig_convergence}
\end{figure}

 We finally show the convergence plot of the global cost function $C(\mathcal{G})$ over the whole GCO process in Fig.~\ref{fig_convergence}. Here the spikes indicate the pruning operations after each iteration, which removes deep branches. The pruning threshold is decreased after two iterations so that more branches are reserved. Although the decrease in cost is extremely subtle after each iteration of pruning, it gradually produces new trees with better global branching structures in the next iteration. Overall, the algorithm reaches convergence after several iterations with spikes due to pruning operations.

\end{document}